\documentclass[notitlepageprb,showpacs,superscriptaddress,preprint,titlepage,aps,pra]{revtex4-2}
\usepackage{amssymb,amsmath,hyperref}
\usepackage{lineno,hyperref,bm,dcolumn,graphicx}

\begin{document}

\title{Enhanced production of $^{8}$Be nuclei in relativistic nuclei fragmentation}

\author{N.G. Peresadko}
\affiliation{Physical Institute P.N. Lebedev RAS, Moscow, Russia,}

\author{A.A. Zaitsev}
\affiliation {Joint Institute for Nuclear Research, Dubna, Russia} 

\author{P.I. Zarubin}
\email{ zarubin@lhe.jinr.ru}
\affiliation {Joint Institute for Nuclear Research, Dubna, Russia} 

\begin{abstract}
	{This article is dedicated to the results of early measuring fragmentation of relativistic $^{16}$O, $^{22}$Ne, $^{28}$Si, and $^{197}$Au nuclei in a nuclear track emulsion. It has been found that there is a contribution of the unstable $^8$Be nucleus decays to $\alpha$-particle multiplicities. These renewed measurements of nuclear track emulsion exposed to $^{84}$Kr nuclei at the energy of 950 MeV per nucleon have been analyzed and given below.}
\end{abstract}

\maketitle

\section{Introduction}

Decays of unstable nuclei $^8$Be $\to$ 2$\alpha$ and $^9$B $\to$ 2$\alpha p$, as well as the Hoyle state HS $\to$ 3$\alpha$ \cite{1,2,3}, can be reconstructed in the events of relativistic nuclei fragmentation in the nuclear track emulsion (NTE). Having extremely low decay energy, the above fragments appear as pairs and triplets with the smallest opening angles. According to their widths, they decay into several thousand ($^8$Be and HS) to several tens ($^9$B) atomic sizes, i.e., by many orders of magnitude longer than the time of appearance of other fragments and can be also considered as fragments. The predicted sizes of these states are unexpectedly large. All these factors make them critical objects for understanding fragmentation, as well as signatures of more complex states of the nuclear-molecular structure which could decay through HS, $^9$B and $^8$Be. It is possible that the unstable states which are part of the nuclear structure reveal themselves in the process of fragmentation. An alternative to the above assumption consists in the $^8$Be formation of the $\alpha$-pair interaction with capturing $\alpha$-particles and nucleons. The consequence of this process would be increasing of the $^8$Be yield with $\alpha$-particle multiplicity $n_{\alpha}$ and, possibly, $^9$B and HS. At the assumption that these states are the nuclear structure we can expect inverse correlation of the $^8$Be yield and $\alpha$-particle multiplicity.

Identification of the above decays requires to reconstruct invariant masses of 2$\alpha$-pairs $Q_{2\alpha}$, 2$\alpha p$-triplets $Q_{2\alpha p}$ and 3$\alpha$-triplets $Q_{3\alpha}$, respectively. In general, the invariant mass $Q = M^* - M$ is given by the sum $M^{*2}$ = $\Sigma(P_i\cdot P_k)$, where $P_{i,k}$ are 4-momenta of fragments, and $M$ is their mass. In the case of relativistic fragmentation in NTE it is feasible to determine $Q$ by means of fragment emission angles assuming that they retain the primary momentum per nucleon, as well as the correspondence of He - $^4$He and H - $^1$H. The selection conditions tested in dissociation of $^9$Be, $^{10}$B, $^{10}$C, $^{11}$C, and $^{12}$C are $Q_{2\alpha}$ ($^8$Be) $\leq$ 0.2 MeV, $Q_{2\alpha p}$($^9$B) $\leq$ 0.5 MeV, and $Q_{3\alpha}$(HS) $\leq$ 0.7 MeV.

\section{Analysis of early exposures}

Recently, the available measurements of interactions $^{16}$O, $^{22}$Ne, $^{28}$Si, and $^{197}$Au nuclei performed by the Emulsion Collaboration at the JINR Synchrophasotron and the EMU Collaboration at AGS (BNL) and SPS (CERN) have been reanalyzed in the approach described above (references in \cite{3}). The search for these nuclear interactions was carried out along primary tracks which enabled us to find different numbers of relativistic fragments of He and H without sampling interactions. These data make it possible to follow the contribution of unstable states and can be used to speed-up searching for events with a higher multiplicity by means of the cross-scan method. 

For $^{16}$O nuclei at 3.65, 14.6, 60, and 200 GeV/nucleon we have observed the increase of this ratio: $n_\alpha$ = 2 (8 $\pm$ 1) to 3 (23 $\pm$ 3) and 4 (46 $\pm$ 14). It also increases for $^{22}$Ne (3.22 GeV/nucleon) at $n_\alpha$ = 2 (6 $\pm$ 1), 3 (19 $\pm$ 3), 4 (31 $\pm$ 6) and $^{28}$Si (14.6 GeV/nucleon) at $n_\alpha$ = 2 (3 $\pm$ 2), 3 (13 $\pm$ 5), 4 (32 $\pm$ 6), 5 (38 $\pm$ 11). The data for $^{197}$Au (10.7 GeV/nucleon) have indicated that the ratio continues to grow strongly towards $n_\alpha$ = 10. Due to increasing multiplicities in the events the measurements in the latter case have become more complicated: that is why we have chosen the following condition $Q_{2\alpha}$($^8$Be) $\leq$ 0.4 MeV. 

The $^{197}$Au interactions contain the triplets $Q_{2\alpha p}$($^9$B) $\leq$ 0.5 MeV and $Q_{3\alpha}$(HS) $\leq$ 0.7 MeV. The ratio of the number of events $N_{n\alpha}$($^9$B), $N_{n\alpha}$(HS), and $N_{n\alpha}$(2$^{8}$Be) to $N_{n\alpha}$($^8$Be) does not show a noticeable change with $n_\alpha$, indicating the increase relative to $N_{n\alpha}$. However, small statistics allows one to characterize only the trend of the unstable states growth. Summing up the statistics $N_{n\alpha}$($^9$B), $N_{n\alpha}$(HS) , and $N_{n\alpha}$(2$^8$Be), over the multiplicity $n_\alpha$ and normalizing it to the sum $N_{n\alpha}$($^8$Be) we have obtained relative contributions equal to 25 $\pm$ 4\%, 6 $\pm$ 2\%, 10 $\pm$ 2\%, respectively.

\section{New measurements}

Statistics of $n_\alpha$ events has been increased by transverse scanning of NTE layers longitudinally exposed to 950 MeV/nucleon $^{84}$Kr nuclei (GSI, early 90s). According to the SRIM program, the deceleration at lengths up to 6 cm is approximately uniform and equal to about 8 MeV/mm (the total range is about 8 cm). Taking into account the effect from vertex positions in the invariant mass makes it possible to use most of the NTE area. In addition, the deceleration increases the fragment emission angles making measurements more convenient. The momentum of the fragments is taken with a factor of 0.8 to approximately count the momentum initial value drop down in the interaction. Being not important to select $Q_{2\alpha}$($^8$Be) $\leq$ 0.4, this correction allowed us to preserve the position of the peak $Q_{3\alpha}$(HS $\to$ $^8$Be$\alpha$).

\begin{figure}
\centering\includegraphics[width=14cm]{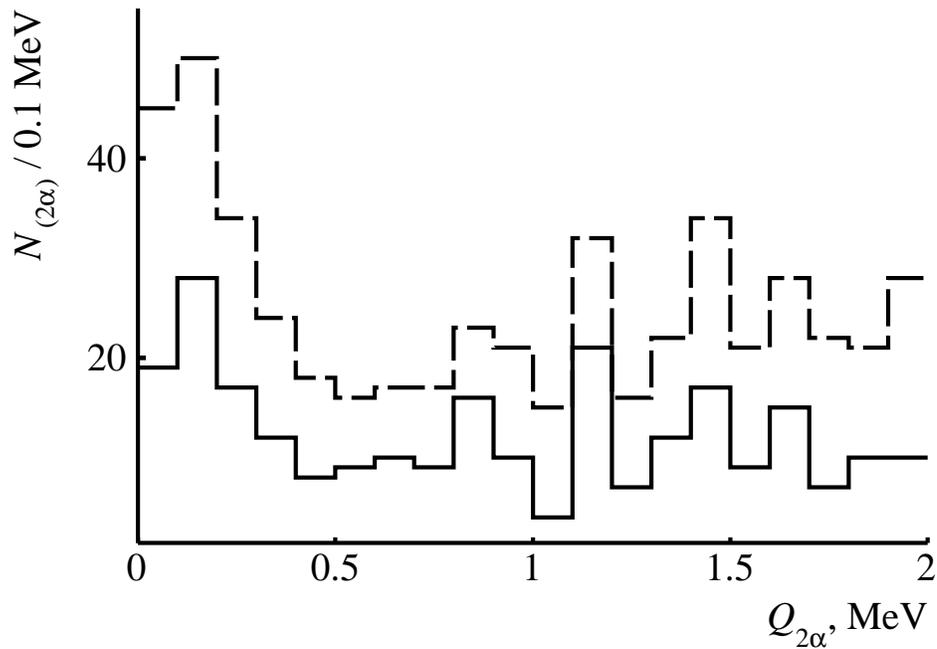}
	\caption{Distribution over invariant mass of $\alpha$-particle pairs $Q_{2\alpha}$ in fragmentation of $^{84}$Kr nuclei at the initial energy of 950 MeV per nucleon in nuclear track emulsion according to the latest measurements (solid line) and early data \cite{4} (added by the dotted line).}
	\label{fig:1}
\end{figure}

Fig. 1 shows the distribution of 85 events $n_\alpha >$ 3 over $Q_{2\alpha}$. The measurements of planar angles in this sample were carried out at the microscope directly by rotating the emulsion layer relative to the primary track. While more efficient, this method is less accurate than the coordinate method. Nevertheless, it is sufficient to select $Q_{2\alpha}$($^8$Be) $\leq$ 0.4 MeV and candidates for more complex decays. In addition, the distributions are supplemented with the data from early measurements of 184 interactions $n_\alpha >$ 2 of $^{84}$Kr at 950-800 MeV per nucleon.  The vertex positions are absent for the above data. Assuming the average energy of 875 MeV/nucleon, this factor turns out to be not critical to identify $Q_{2\alpha}$($^8$Be) $\leq$ 0.4 MeV. The ratios $N_{n\alpha}$($^8$Be) and $N_{n\alpha}$ (\%) for the both samples are $n_\alpha$ = 4 (24 $\pm$ 6), 5 (27 $\pm$ 6), 6 (53 $\pm$ 15) and the sum $n_\alpha >$ 6 (64 $\pm$ 14). The new sample contains the 2$^8$Be event at $n_{\alpha}$ = 6 isolated in the initial part of the $Q_{4\alpha}$ spectrum at 0.6 MeV. 

Thus, the universal effect of increasing the probability to detect $^8$Be in any event having the growth in the multiplicity of $\alpha$-particles in it, has been confirmed for one more nucleus. The presented data are the first contribution to the targeted search for decays of 4$\alpha$-particle states of the Bose-Einstein condensate in the events of relativistic fragmentation of heavy nuclei in the nuclear emulsion based on the most accurate measurements of coordinates over the tracks.

\end{document}